\documentclass[
aps,physrev,
reprint,
superscriptaddress,nofootinbib,amsmath,dvipsnames
]{revtex4-2}

\usepackage{orcidlink}
\usepackage{booktabs}
\usepackage{aas_macros}

\usepackage[normalem]{ulem}

\hypersetup{
  colorlinks=true,
  allcolors=Blue
}

\begin{document}


\title{Eccentricity distribution of extreme mass ratio inspirals}

\author{Davide Mancieri\,\orcidlink{0009-0004-5106-9363}}
\email[Contact author: ]{d.mancieri@campus.unimib.it}
\affiliation{
 Dipartimento di Fisica, Università degli Studi di Trento, Via Sommarive 14, 38123 Povo, Italy
}
\affiliation{
 Dipartimento di Fisica ``G. Occhialini'', Università degli Studi di Milano-Bicocca, Piazza della Scienza 3, 20126 Milano, Italy
}
\affiliation{
 INFN - Sezione di Milano-Bicocca, Piazza della Scienza 3, 20126 Milano, Italy
}

\author{Luca Broggi\,\orcidlink{0000-0002-9076-1094}}
\affiliation{
 Dipartimento di Fisica ``G. Occhialini'', Università degli Studi di Milano-Bicocca, Piazza della Scienza 3, 20126 Milano, Italy
}
\affiliation{
 INFN - Sezione di Milano-Bicocca, Piazza della Scienza 3, 20126 Milano, Italy
}

\author{Morgan Vinciguerra\,\orcidlink{0009-0009-2589-581X}}
\affiliation{
 Dipartimento di Fisica ``G. Occhialini'', Università degli Studi di Milano-Bicocca, Piazza della Scienza 3, 20126 Milano, Italy
}

\author{Alberto Sesana\,\orcidlink{0000-0003-4961-1606}}
\affiliation{
 Dipartimento di Fisica ``G. Occhialini'', Università degli Studi di Milano-Bicocca, Piazza della Scienza 3, 20126 Milano, Italy
}
\affiliation{
 INFN - Sezione di Milano-Bicocca, Piazza della Scienza 3, 20126 Milano, Italy
}
\affiliation{
 INAF - Osservatorio Astronomico di Brera, Via Brera 20, 20121 Milano, Italy
}

\author{Matteo Bonetti\,\orcidlink{0000-0001-7889-6810}\,}
\affiliation{
 Dipartimento di Fisica ``G. Occhialini'', Università degli Studi di Milano-Bicocca, Piazza della Scienza 3, 20126 Milano, Italy
}
\affiliation{
 INFN - Sezione di Milano-Bicocca, Piazza della Scienza 3, 20126 Milano, Italy
}
\affiliation{
 INAF - Osservatorio Astronomico di Brera, Via Brera 20, 20121 Milano, Italy
}

\date{\today}

\begin{abstract}
We present realistic eccentricity distributions for extreme mass ratio inspirals (EMRIs) forming via the two-body relaxation channel in nuclear star clusters, tracking their evolution up to the final plunge onto the central Schwarzschild massive black hole (MBH). We find that EMRIs can retain significant eccentricities at plunge, with a distribution peaking at $e_\mathrm{pl} \approx0.2$, and a considerable fraction reaching much higher values. In particular, up to $20\%$ of the forming EMRIs feature $e_\mathrm{pl} > 0.5$ for central MBH masses $M_\bullet$ in the range $10^5 \, \mathrm{M_\odot} \leq M_\bullet \leq 10^6 \, \mathrm{M_\odot}$, partially due to EMRIs forming at large semi-major axes and ``cliffhanger EMRI'', usually neglected in literature. This highlights the importance of accounting for eccentricity in waveform modeling and detection strategies for future space-based gravitational wave observatories such as the upcoming Laser Interferometer Space Antenna (LISA). Furthermore, we find that the numerical fluxes in energy and angular momentum currently implemented in the FastEMRIWaveforms (FEW) package may not adequately sample the full parameter space relevant to low-mass MBHs ($M_\bullet < 10^6 \, \mathrm{M_\odot}$), potentially limiting its predictive power in that regime. Specifically, for $M_\bullet=10^5 \, \mathrm{M_\odot}$ we find that about $75\%$ ($50 \%$) of EMRIs at 2 years (6 months) from plunge fall outside the currently available flux parameter space. Our findings motivate the development of extended flux grids and improved interpolation schemes to enable accurate modeling of EMRIs across a broader range of system parameters.
\end{abstract}

\maketitle

\section{Introduction}
Massive black holes (MBHs) with masses in the range of $10^6 \,\text{-}\, 10^{10} \, \mathrm{M_\odot}$ are commonly found at the cores of most galaxies \citep{1995ARA&A..33..581K,1998AJ....115.2285M,2013ARA&A..51..511K}. Lighter MBH candidates with masses of $10^4 \,\text{-} \,10^5 \, \mathrm{M_\odot}$ have also been identified in dwarf galaxies \citep{2020ARA&A..58..257G}. These MBHs typically reside in dense stellar environments known as nuclear star clusters \citep{1997AJ....114.2366C,2002AJ....123.1389B,2006ApJS..165...57C,2007A&A...469..125S,2009A&A...502...91S,2012ApJS..203....5T}, where frequent dynamical interactions occur \citep{2008gady.book.....B,2013degn.book.....M,2017ARA&A..55...17A}. Evidence of such interactions arises in the electromagnetic spectrum due to extreme phenomena like tidal disruption events involving stars and MBHs (see Refs.\ \citep{2021ARA&A..59...21G,2021TDE} for recent reviews).

The upcoming Laser Interferometer Space Antenna (LISA, see Refs.\ \citep{2017arXiv170200786A,2024arXiv240207571C}) will open a new observational window onto galactic nuclear interactions by detecting gravitational waves (GWs, see Ref.\ \citep{2007gwte.book.....M}) from compact objects, likely stellar-mass black holes (BHs), spiraling into MBHs. These systems, known as extreme mass ratio inspirals (EMRIs, see Refs.\ \citep{2007CQGra..24R.113A,2018LRR....21....4A,2022hgwa.bookE..17A} for reviews) due to typical mass ratios in the range $10^{-5}\,\text{-}\,10^{-3}$, offer a unique probe of strong-field gravity.

EMRI detections will yield insights in astrophysics \citep{2023LRR....26....2A}, cosmology \citep{2023LRR....26....5A}, fundamental physics \citep{2022LRR....25....4A}, and waveform modeling \citep{2023arXiv231101300L}. Indeed, EMRIs are predicted to spend years in LISA band, accumulating high signal-to-noise ratios over around $10^4 \,\text{-}\, 10^5$ orbital cycles \citep{2004PhRvD..69h2005B}. This will enable extremely precise measurements of MBH masses and spins \citep{2017PhRvD..95j3012B} for the otherwise quiescent MBH population not emitting in the electromagnetic spectrum \citep{2010PhRvD..81j4014G,2019ApJ...883L..18G}. The long-lived GW signal also makes EMRIs sensitive to environmental effects \citep{2014PhRvD..89j4059B}: the presence of matter near the MBH might cause slight phase shifts in the waveform, allowing LISA to test for accretion disks \citep{2011PhRvD..84b4032K,2011PhRvL.107q1103Y,2014PhRvD..89j4059B,2025PhRvD.111h4006D} and dark matter structures \citep{2024PhRvL.133l1404D}.

EMRIs have also been proposed as sources of quasi-periodic eruptions (QPEs, \emph{e.g.}\ Ref.\ \citep{2023A&A...675A.100F}), currently unexplained soft X-ray flares repeating over few hours to few days, coming from the nuclei of some galaxies \citep{2019Natur.573..381M,2020A&A...636L...2G,2021Natur.592..704A,2024A&A...684A..64A,2025arXiv250617138A,2021ApJ...921L..40C,2025ApJ...983L..39C,2023A&A...675A.152Q,2024Natur.634..804N,2025MNRAS.540...30B,2025NatAs...9..895H}. This could enable multi-messenger observations of EMRIs in the future, although current QPE detections seem to point to precursors of LISA detectable EMRIs orbiting at lower frequencies \citep{2025arXiv250510488S}.

To fully exploit EMRI detections with LISA, we need fast and accurate waveform models for them. This has motivated the development of the FastEMRIWaveforms (FEW, see Refs.\ \citep{2021PhRvL.126e1102C,2021PhRvD.104f4047K,2023arXiv230712585S,2025arXiv250609470C,chapman_bird_2025_15630565}) package, which is widely adopted in the community. FEW can generate fully relativistic, adiabatic waveforms for equatorial orbits around Kerr MBHs in under a second, although this is currently limited to eccentricities below 0.9.

Whether this is a limitation or not for EMRI scientific exploitation, it depends on the considered formation channel. For example, this is not expected to be an issue for EMRIs forming in accretion disks \citep{2007MNRAS.374..515L,2021PhRvD.103j3018P,2023MNRAS.521.4522D}, where gas interactions circularize the orbit, or for EMRIs forming from the tidal separation of stellar binaries \citep{1988Natur.331..687H,2005ApJ...631L.117M} which also tend to be only moderately eccentric. However, other formation channels might yield highly eccentric EMRIs within the LISA band. For instance, the Lidov–Kozai mechanism \citep{1910AN....183..345V,1962P&SS....9..719L,1962AJ.....67..591K}, triggered by a secondary MBH, can drive large eccentricity oscillations on the orbit of the smaller BH \citep{2011ApJ...729...13C,2014MNRAS.438..573B,2022ApJ...927L..18N,2022MNRAS.516.1959M}. EMRIs may also form following supernova kicks \citep{2017MNRAS.469.1510B,2019MNRAS.485.2125B}, again leading to high eccentricities.

Although all of the aforementioned channels might be at work, significantly contributing to the LISA EMRI rates, this latter is expected to be dominated by what is considered the ``standard'' EMRI formation mechanism: two body relaxation in dense nuclei (see Ref.\ \citep{2018LRR....21....4A} and references therein). In short, objects orbiting an MBH in a nuclear star cluster undergo frequent orbital deflections through weak two-body encounters with other bodies, which are themselves subject to the same interactions. This relaxation process is mainly driven by angular momentum exchanges, which randomly modify the orbital eccentricity until the orbit reaches a sufficiently small pericenter for GWs emission to dominate over two-body relaxation, thus triggering the formation of an EMRI.
In this work, we focus on this mechanism and we derive the expected EMRI eccentricity distribution for Schwarzschild MBHs, to verify whether current waveform models meet the needs of what is considered to be the dominant EMRI formation channel.

Specifically, we build upon the results of \citet{2025A&A...694A.272M}, who modeled EMRI formation using a Monte Carlo scheme to consistently account for perturbations acting on the 2.5 post-Newtonian (PN, see Ref.\ \citep{2024LRR....27....2S}) orbital motion of a single stellar-mass BH due to weak encounters with other objects surrounding the central MBH. That study focused on EMRI formation and thus stopped simulations before the final inspiral. Here, we take their endpoint binary configurations and evolve them to plunge using FEW, to then apply appropriate weights to recover the underlying astrophysical eccentricity distribution.

The paper is organized as follows. In Sect.\ \ref{sec:methods} we present the methods we employ to integrate and then weight EMRIs; in Sect.\ \ref{sec:results} we present and discuss our results; in Sect.\ \ref{sec:comp} we compare our findings with previous works; in Sect.\ \ref{sec:conc} we draw our conclusions. Throughout this work, we adopt $M_\bullet$ for the MBH mass, $G$ for the gravitational constant, and $c$ for the speed of light in vacuum.

\section{Methods} \label{sec:methods}
    In this section, we describe the methods we use to extract the eccentricity distribution of the EMRI population generated by \citet{2025A&A...694A.272M} at various stages of their inspiral. We first detail the evolution scheme we employ to integrate this EMRI population up to the final plunge, then the weighting procedure we follow to produce astrophysically consistent eccentricity distributions.
    
\subsection{EMRI trajectory integration} \label{sec:int}
\begin{figure}
        \centering
        \resizebox{\hsize}{!}{\includegraphics{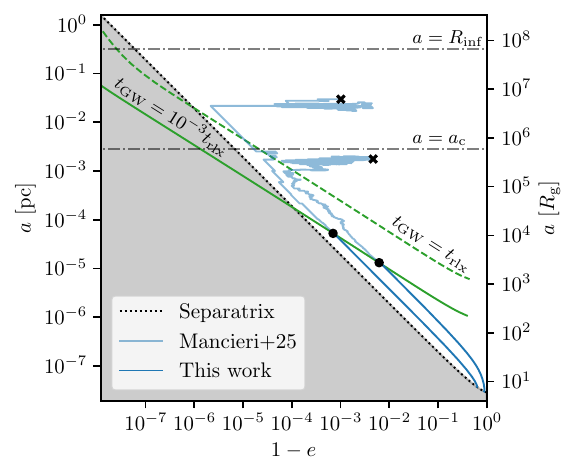}}
        \caption{Two example EMRIs forming and plunging into a $10^5 \, \mathrm{M_\odot}$ MBH. The early part of the integration (blue shaded tracks) is from \citet{2025A&A...694A.272M}, while the final phase (blue solid tracks) is presented in this work. The gray dotted line marks the separatrix, and the gray shaded region below it is the ``relativistic'' loss cone. The loss cone is usually delimited by the pericenter condition $r_\mathrm{p} = 8 R_\mathrm{g}$, while here the separatrix corresponds to $r_\mathrm{p} = 4 R_\mathrm{g}$ for $e \to 1$ (see Sect.\ \ref{sec:rates}). The green dashed curve separates the region dominated by two-body relaxation (above) from that dominated by GWs emission (below). Black crosses mark the initial conditions from \citet{2025A&A...694A.272M}, as well as the initial semi-major axes $a_\mathrm{i}$ used in this work to weight EMRIs (see Sect.\ \ref{sec:rates}). Black dots indicate the stopping points of runs in \citet{2025A&A...694A.272M}, which also serve as initial conditions in this work. Both an initially wide ($a_\mathrm{i} > a_\mathrm{c}$) and a classical ($a_\mathrm{i} < a_\mathrm{c}$) EMRI are shown (see Sect.\ \ref{sec:results}).}
        \label{fig:veryniceplot}
    \end{figure}

    In the simulations by \citet{2025A&A...694A.272M}, EMRIs were identified, and their integration stopped, as soon as their GW emission timescale $t_\mathrm{GW}=10^{-3}\,t_\mathrm{rlx}$, where $t_\mathrm{rlx}$ is the angular momentum relaxation time of the orbit (see Fig.\ \ref{fig:veryniceplot})\footnote{In the case of $M_\bullet = 4 \times 10^6 \, \mathrm{M_\odot}$, a significant fraction of EMRIs were stopped at $t_\mathrm{GW} < t_\mathrm{rlx} = 10 \, \mathrm{Myr}$, before the condition $t_\mathrm{GW} = 10^{-3} t_\mathrm{rlx}$ was reached. A similar correction was also applied to lower MBH masses, but it impacted a much smaller number of EMRIs.}. From this point onward, it is safe to assume that the dynamics is completely decoupled from two-body relaxation and determined by GWs emission only.

    From the output of the simulations by \citet{2025A&A...694A.272M}, we extract the final 2PN orbital energy (not including the stellar potential) and angular momentum $L$ per unit of the reduced mass of the binary (the mass of the secondary is fixed to $10 \, \mathrm{M_\odot}$ throughout this work). We then find the energy $E$ of the corresponding geodesic by adding $c^2$ to the 2PN estimate to account for the rest energy of the secondary\footnote{A more accurate conversion would be $E_\mathrm{GR}=c^2\sqrt{2E_\mathrm{Newt}/c^2+1}$, but it expands to $E_\mathrm{GR} \approx E_\mathrm{Newt}+c^2$ for small $E_\mathrm{Newt}/c^2$. This is always the case for the highly eccentric orbits we are considering.} \citep{2017grav.book.....M}. $E$ and $L$ are then converted into the the initial $(p_0, e_0)$ that we wish to integrate up to plunge, where $p$ is the semi-latus rectum and $e$ is the eccentricity of the orbit. We verify that this conversion returns sensible values, compatible to approximate estimates obtained considering the final pericenter and apocenter passages of the original simulations in \citet{2025A&A...694A.272M}. To perform the change of parameters we use the \texttt{ELQ\_to\_pex} function provided by the FEW package, which implements the $(E,L_\mathrm{z},Q)\to (p,e,x_\mathrm{I})$ map developed by \citet{2024PhRvD.109f4077H}. Here $Q$ is the Carter constant \citep{1968PhRv..174.1559C}, while $x_\mathrm{I}$ is the cosine of the maximum inclination angle between the orbital and the MBH spin planes (see \emph{e.g.}\ Ref.\ \citep{2024arXiv241104955L}), measured in Boyer-Lindquist coordinates \citep{1967JMP.....8..265B}. As we consider Schwarzschild MBHs in this work, $Q=0$ and $x_\mathrm{I}=1$. For the same reason, the angular momentum along the MBH spin direction $L_\mathrm{z}$ is simply the same as $L$.

    Since orbits in general relativity are not ellipses, $p$ and $e$ are not to be interpreted as geometrical parameters, but simply as functions of the pericenter $r_\mathrm{p}$ and apocenter $r_\mathrm{a}$ of the orbit:
    \begin{align}
        p = \frac{2 r_\mathrm{a} r_\mathrm{p}}{r_\mathrm{a}+r_\mathrm{p}} \, ,  \quad   e = \frac{r_\mathrm{a} - r_\mathrm{p}}{r_\mathrm{a}+r_\mathrm{p}} \, .
    \end{align}
    In turn, $r_\mathrm{p}$ and $r_\mathrm{a}$ are the two largest roots of the radial effective potential of the MBH, given $E$ and $L$ (\emph{e.g.}\ Refs.\ \citep{2002CQGra..19.2743S,2002PhRvD..66d4002G,2009CQGra..26m5002F,2020CQGra..37n5007V}).

\begin{figure}
        \centering
        \resizebox{\hsize}{!}{\includegraphics{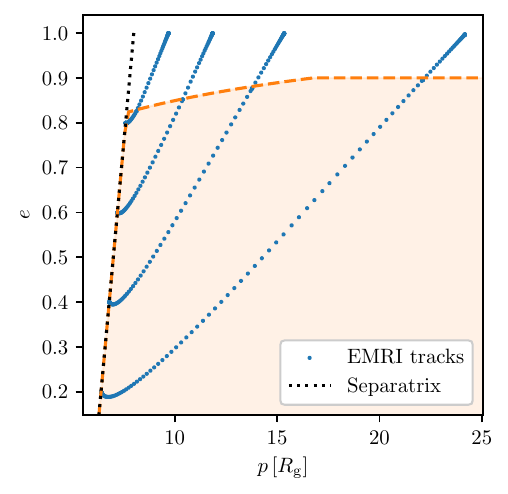}}
        \caption{Integration of some of our initial conditions with the method described in Sect.\ \ref{sec:int}. The blue dots show the steps taken, while the gray dotted line is the separatrix. The orange shaded region displays the range covered by the precomputed flux grids, inside which the \texttt{KerrEccEqFlux} model can be used. We switch to this model from the \texttt{PN5} one when crossing the dashed orange curve.}
        \label{fig:example}
    \end{figure}

    To integrate ($p,e$) down to plunge from their initial values ($p_0,e_0$), we employ the trajectory package within FEW. The latest version of FEW includes two main trajectory modules \citep{2025arXiv250609470C}:
    \begin{enumerate}
        \item \texttt{KerrEccEqFlux}. The module is based on numerical computation of time-averaged fluxes in energy and angular momentum lost by the system (both to infinity and through the event horizon of the primary) due to GWs radiation. The fluxes are computed from the Teukolsky equation \citep{1973ApJ...185..635T} using BH perturbation theory techniques (\emph{e.g.}\ Refs.\ \citep{2020PhRvD.102f4005F,2021PhRvD.103j4014H,2021PhRvD.103j4045S}). The computation of the fluxes is highly expensive, and thus it is performed offline for some discrete values of the orbital parameters. During the integration, this trajectory module simply interpolates over the precomputed grid of fluxes. As a result, the module can only be applied to pairs $(p,e)$ within the precomputed range. In the case of Schwarzschild MBHs, for $p \gtrsim 16.8 \, R_\mathrm{g}$, this limits the eccentricity to $e < 0.9$, where $R_\mathrm{g}=GM_\bullet/c^2$ is the gravitational radius of the MBH. For lower values of the semi-latus rectum, the eccentricity is further limited, dropping to  $e \lesssim 0.82$ for the smallest allowed $p$ (see Fig.\ \ref{fig:example}).
        \item \texttt{PN5}. The module is based on analytical expressions for the inspiraling force, valid to $\mathcal{O} \big( (v/c)^{11}, e^{11} \big)$ order \citep{2020PhRvD.102f4005F}. Here $v/c=\sqrt{R_\mathrm{g}/p}$ is the small parameter over which the PN expansion is performed. This module can be applied at any $(p,e)$, even though its accuracy decreases for small $p$ and large $e$.
    \end{enumerate}
    Integration performed with these modules is accurate at adiabatic order, meaning we are under the assumption that the orbital elements evolve slowly.

    Our initial values $e_0$ (roughly ranging from 0.99 to 0.9997) always fall outside the domain of validity of the \texttt{KerrEccEqFlux} trajectory model. We therefore proceed with the integration in two steps: \emph{(i)} we integrate with the \texttt{PN5} module to bring $(p,e)$ inside the range of the precomputed flux grids;  \emph{(ii)} we switch to the \texttt{KerrEccEqFlux} module to get the system to the plunge. In some cases, the plunge occurs before it is possible to switch module, so that the whole integration is performed with the \texttt{PN5} module only. This is always the case for EMRIs that plunge at $e_\mathrm{pl} \gtrsim 0.82$ (see Fig.\ \ref{fig:example}). The plunging condition is identified by $p = p_\mathrm{sep}$, where $p_\mathrm{sep}$ is the semi-latus rectum of the innermost bound stable orbit, also known as separatrix. For a Schwarzschild MBH\footnote{To mitigate numerical issues that might arise using the full separatrix expression for generic orbits around Kerr MBHs, we perform the integration enabling the \texttt{enforce\_schwarz\_sep} flag in the trajectory class. In typical runs the integration is terminated once $p/R_\mathrm{g} -6-2e = 2 \times10^{-3}$. In case the integration is performed with the \texttt{PN5} module only, it instead stops at $p/R_\mathrm{g} -6-2e = 10^{-1}$. This affects only few systems and only the high-eccentricity tail ($e_\mathrm{pl}\gtrsim 0.82$) of Fig.\ \ref{fig:plunge_distr}.} this is given by \citep{2020PhRvD.101f4007S}
    \begin{equation} \label{eq:sep}
        p_\mathrm{sep} = ( 6 + 2e ) R_\mathrm{g} \, .
    \end{equation}
    We repeat the integration for all initial conditions twice: \emph{(i)} by directly integrating the orbital elements $(p,e)$, and \emph{(ii)} by exploiting the constants of motion $(E,L)$. The latter choice is more numerically robust near the separatrix \citep{2024PhRvD.109f4077H}, although issues can arise for almost circular orbits, where small errors in the constants of motion can be amplified when transformed back to $(p,e)$ within FEW. For this work, we opted for the integration of $(E,L)$, as we found that both methods give almost indistinguishable results at the precision level we require.

    Figure \ref{fig:example} shows a selection of the EMRI evolution tracks that we produce.

\subsection{Astrophysical formation rates}
    \label{sec:rates}
        \begin{figure}
        \centering
        \resizebox{\hsize}{!}{\includegraphics{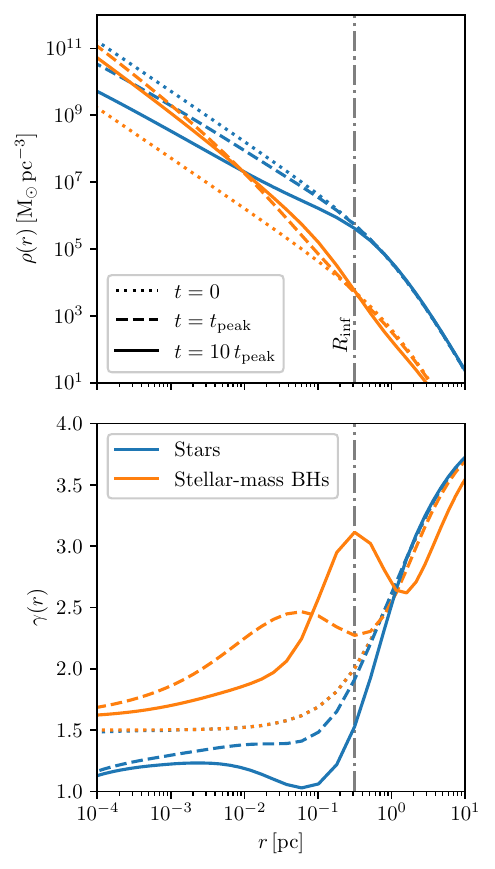}}
        \caption{In the upper panel, time dependent density profiles for stars (blue) and stellar-mass BHs (orange) in our nuclear star cluster model with an MBH of $10^5 \, \mathrm{M_\odot}$. Both components are initially distributed with an inner slope of 1.5 but different normalizations. Profiles are shown at three different times: the initial condition (dotted lines), $t_\mathrm{peak}$ (dashed lines), and $10 \, t_\mathrm{peak}$ (solid lines). In the lower panel, time evolution of the local slope of the density profiles as a function of radius $r$.}
        \label{fig:new_fig}
    \end{figure}
    We employ loss cone theory to weight EMRIs based on their semi-major axis $a$ at the start of the runs performed in \citet{2025A&A...694A.272M} (see Fig.\ \ref{fig:veryniceplot}), depending on their formation rate in the nuclear star cluster. As in \citet{2025A&A...694A.272M}, we solve the orbit averaged Fokker-Planck equation to compute the time evolution of the distribution function of a spherical stellar system because of two-body relaxation \citep{1978ApJ...226.1087C}.
    
    Loss cone theory provides a framework to compute the rate of particles pushed at pericenter smaller than the loss cone radius $R_\mathrm{lc} = 8 \, R_\mathrm{g}$. Considering that the Fokker-Planck coefficients are computed in Newtonian dynamics, the value of $R_\mathrm{lc}$ is chosen so that a highly eccentric orbit with such a pericenter in the potential $V_\mathrm{Newt}(r) = -GM_\bullet/r +  L^2/2r^2$ has $L=4\,c\,R_\mathrm{g}$, the limiting value prescribed by general relativity for a particle to avoid plunging on a Schwarzschild MBH \citep{1994PhRvD..50.3816C,2011PhRvD..84d4024M,2025A&A...694A.272M}.
    
    Because of the stellar potential, we label the specific energy in the potential of the nuclear star cluster $\mathcal{E}$, and infer the semi-major axis $a(\mathcal{E})$ from the radius of the circular orbit $r_\mathrm{c}$ at that energy, using the same mapping as in \citet{2025A&A...694A.272M}. The rate $\mathrm{d}\dot{N}_\mathrm{lc}$ of objects captured with energy between $\mathcal{E}$ and $\mathcal{E}+\mathrm{d}\mathcal{E}$ can be expressed in terms of the differential rate $\mathcal{F}_\mathrm{lc}$, also known as the loss cone flux:
    \begin{equation}
        \mathrm{d}\dot{N}_\mathrm{lc}(t,\mathcal{E}) = \mathcal{F}_\mathrm{lc}(t,\mathcal{E}) \, \mathrm{d} \mathcal{E}\, ,
    \end{equation}
    where $\mathcal{F}_\mathrm{lc}$ can be computed directly from the distribution function. As shown in \citet{2022MNRAS.514.3270B}, $\mathcal{F}_\mathrm{lc}$ is highly time-dependent, reflecting the impact of mass segregation on the evolving distribution of objects in the cluster. Moreover, $\mathcal{F}_\mathrm{lc}(t,\mathcal{E})$ goes to zero for $\mathcal{E}\to0$ and $\mathcal{E}\to\infty$, and has a peak around $a = 0.1\,\text{-}\,1\, R_\mathrm{inf}$. Here $R_\mathrm{inf}$ is the influence radius of the central MBH, which we define as $R_\mathrm{inf} = G M_\bullet/\sigma^2_\mathrm{inf}$ \citep{1972ApJ...178..371P}, where $\sigma_\mathrm{inf} = \sigma(R_\mathrm{inf})$ is the velocity dispersion of the stellar distribution measured at the influence radius. Both in \citet{2025A&A...694A.272M} and in this work, $\sigma_\mathrm{inf}$ is related to the mass of the central MBH through the M--$\sigma$ relation \citep{2000ApJ...539L...9F,2000ApJ...539L..13G}, according to the best fit from \citet{2009ApJ...698..198G}.
    
    Not all captured orbits result in EMRIs, as they can also produce plunges. Since we are interested in the former, it is necessary to multiply the loss cone flux by a function $S(\mathcal{E})$ that quantifies, at any given energy, the fraction of captures that are EMRIs. The differential rate of EMRIs is therefore
    \begin{equation}
        \mathcal{F}_\mathrm{EMRI}(t,\mathcal{E}) = \mathcal{F}_\mathrm{lc}(t,\mathcal{E})\,S(\mathcal{E}) \, .
    \end{equation}
    $S(\mathcal{E})$ is always $1$ for $a\to0$; at larger $a$ it decreases to a lower value, which is zero for $M_\bullet \gtrsim 3\times10^5 \, \mathrm{M_\odot}$ \citep{2005ApJ...629..362H,2021MNRAS.501.5012R}, but larger than zero for lower central MBH masses \citep{2024PhRvL.133n1401Q,2025A&A...694A.272M}. In particular, we use the numerical $S(\mathcal{E})$ presented in \citet{2025A&A...694A.272M}\footnote{See Sect.\ 4.4 and specifically Eq.\ 61 in \citet{2025A&A...694A.272M}.}.

    We simulate the relaxation process of stars and stellar-mass BHs within nuclear star clusters surrounding MBHs with mass $M_\bullet\in$\{$10^4$, $10^5$, $3\times10^5$, $10^6$, $4\times10^6$\} $\mathrm{M}_\odot$ with the software presented in \citet{2022MNRAS.514.3270B}. The stellar distribution is characterized by two components: \emph{(i)} stars of $1 \,\mathrm{M}_\odot$ each, for a total stellar mass of $20 M_\bullet$; \emph{(ii)} stellar-mass BHs of $10 \,\mathrm{M}_\odot$ each, for a total mass in BHs of $0.2 M_\bullet$. We initialize both components on the same Dehnen profile \citep{1993MNRAS.265..250D,1994AJ....107..634T} with inner slope $1.5$ and scale radius $4\,R_\mathrm{inf}$. The initial slopes align with values commonly employed in the literature (\emph{e.g.}\ Refs.\ \citep{1977ApJ...216..883B,2022MNRAS.514.3270B}), and are consistently evolved with time following mass segregation. The scale radius of the distribution is chosen such that the stellar cluster surrounding our $4 \times 10^6 \, \mathrm{M_\odot}$ matches the observed density at the influence radius of SgrA$^*$ (\emph{e.g.}\ Ref.\ \citep{2007A&A...469..125S}). For each system, we compute $\mathcal{F}_\mathrm{EMRI}$ as a function of time and stop the simulation at $t_\mathrm{end}$, which is $10$ times the time-to-peak $t_\mathrm{peak}$ of the EMRI rate\footnote{In the case of $M_\bullet=4\times10^6 \, \mathrm{M_\odot}$, we stop at $t_\mathrm{end}=10\, \mathrm{Gyr}$, since $t_\mathrm{peak} \approx2\, \mathrm{Gyr}$.}. Indeed, as $\mathcal{F}_\mathrm{lc}$ depends on time, so does $\mathcal{F}_\mathrm{EMRI}$: \citet{2022MNRAS.514.3270B} showed that the EMRI rates reach a maximum at $t_\mathrm{peak}$ and then undergo a nearly exponential decay. Moreover, the maximum was found to be in agreement steady state EMRI rate values quoted in the literature.

    Figure \ref{fig:new_fig} shows the time evolution of the density profiles for stars and stellar-mass BHs. Due to mass segregation, stars are displaced toward larger radii, while stellar-mass BHs initially sink toward the central MBH, resulting in a peak of EMRI formation at $t_\mathrm{peak}$. At later times, as the cluster continues to relax and expand, the BHs are also redistributed to larger radii. Consequently, the stellar density profile becomes progressively shallower in the inner region, whereas the BH profile first steepens and then slowly flattens.

    \begin{figure}
        \centering
        \resizebox{\hsize}{!}{\includegraphics{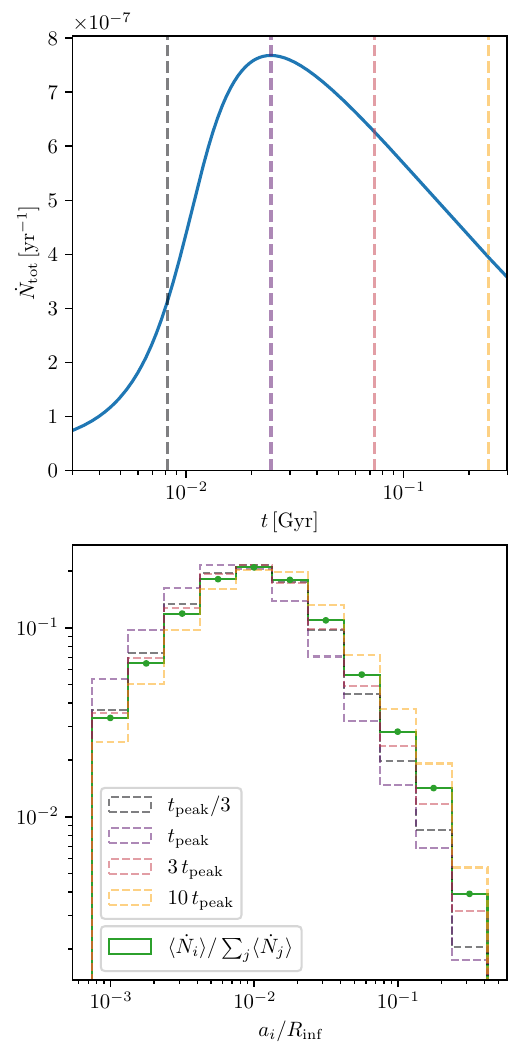}}
        \caption{
        In the upper panel, time dependent formation rate of EMRIs $\dot{N}_\mathrm{tot}$ as a function of time for $M_\bullet=10^5\, \mathrm{M_\odot}$. $\dot{N}_\mathrm{tot}$ is computed at each time as the sum of $\dot{N}_i (t)$ over all energy cells. Vertical dashed lines correspond to $t/t_\mathrm{peak} = 1/3$ (gray), $1$ (purple), $3$ (pink), and $10$ (yellow). In the lower panel, we show (solid green histogram) the average relative number $\langle \dot{N}_i\rangle / \sum_j \langle\dot{N}_j\rangle$ of EMRIs produced at semi-major axis $a_i$ (green dots). We also plot the relative number $\dot{N}_i(t) / \sum_j \dot{N}_j(t)$ at the times marked in the upper panel (dashed histograms, same colors as above). The relative number has a similar shape at all times, but its peak semi-major axis changes according to two-body relaxation: it first decreases because of mass segregation, and later increases because of diffusion.}
        \label{fig:rates}
    \end{figure}

    We introduce a log uniform distribution in semi-major axis centered at the values $a_\mathrm{i}$ used as initial conditions in \citet{2025A&A...694A.272M}. This grid in $a$ is then mapped to energy space using the energy of the corresponding circular orbit. We compute the average rate of EMRIs produced in the $i$-th energy cell as
    \begin{equation}
        \left \langle \dot{N}_i \right \rangle = \frac{1}{t_\mathrm{end}} \int_0^{t_\mathrm{end}} \mathrm{d}t \, \dot{N}_i (t)\, ,
    \end{equation}
    with\footnote{The integral extends from $\mathcal{E}_\mathrm{low}$ to $\mathcal{E}_\mathrm{up}$, such that $\log \big(r_\mathrm{c}(\mathcal{E}_\mathrm{low})\big) = \log(a_i) - \delta/2$ and $\log \big(r_\mathrm{c}(\mathcal{E}_\mathrm{up})\big) = \log(a_i) + \delta/2$, where $\delta$ is the step of the logarithmic grid in $a$.}
    \begin{equation}
        \dot{N}_i (t) =\int_{i-\mathrm{th \ cell}} \mathrm{d}\mathcal{E}\;\mathcal{F}_\mathrm{EMRI}(t, \mathcal{E}) \, .
    \end{equation}
    
    Physically, by time-averaging the rate we are averaging over an ensemble of nuclear star clusters, assuming each is observed at a random stage of relaxation. We find that computing weights using $\langle \dot{N}_i\rangle$ or $\dot{N}_i(t)$ at a given time does not impact qualitatively our results, thus we use the average rate in the rest of this work.
    
    In Fig.~\ref{fig:rates} we consider the case $M_\bullet=10^5 \, M_\odot$ and compare the fraction of EMRIs in the sample that form with semi-major axis $a_i$ at time $t = \{1/3,\ 1,\ 3,\ 10\}\,t_\mathrm{peak}$, together with the time average. At later times larger values of $a_i$ are increasingly favored. In fact, after the initial migration of BHs towards the central region because of mass segregation, at time $t \simeq t_\mathrm{peak}$ the system enters a phase of expansion driven by diffusion. This shifts the the peak of $\mathcal{F}_\mathrm{lc}(t,\mathcal{E})$ to orbits with larger and larger semi-major axes as relaxation proceeds \citep{2022MNRAS.514.3270B}.

    Finally, for a chosen value of $M_\bullet$, we weight each of the $n_i$ EMRIs formed from the initial semi-major axis $a_i$ as
    \begin{equation}
        w_i = \frac{1}{n_i}\times \frac{\langle\dot{N}\rangle_i}{\sum_{j} \langle \dot{N}_j\rangle}
    \end{equation}
    to compute the astrophysically informed distribution of EMRI parameters.

    In Appendix \ref{app:alternative} we also present an alternative weighting procedure, which is based on the common practice of sharply identifying EMRI formation versus direct plunging depending on the semi-major axis of the orbit, instead of slowly transitioning between the two regimes using the smooth transfer function $S$. We show that this choice affects the eccentricity distribution of EMRIs at plunge, but the resulting distributions are qualitatively similar to those obtained with our more accurate procedure.

\section{Results: EMRI eccentricity distributions}\label{sec:results}
\begin{figure*}
            \centering
            \resizebox{0.775\hsize}{!}{\includegraphics{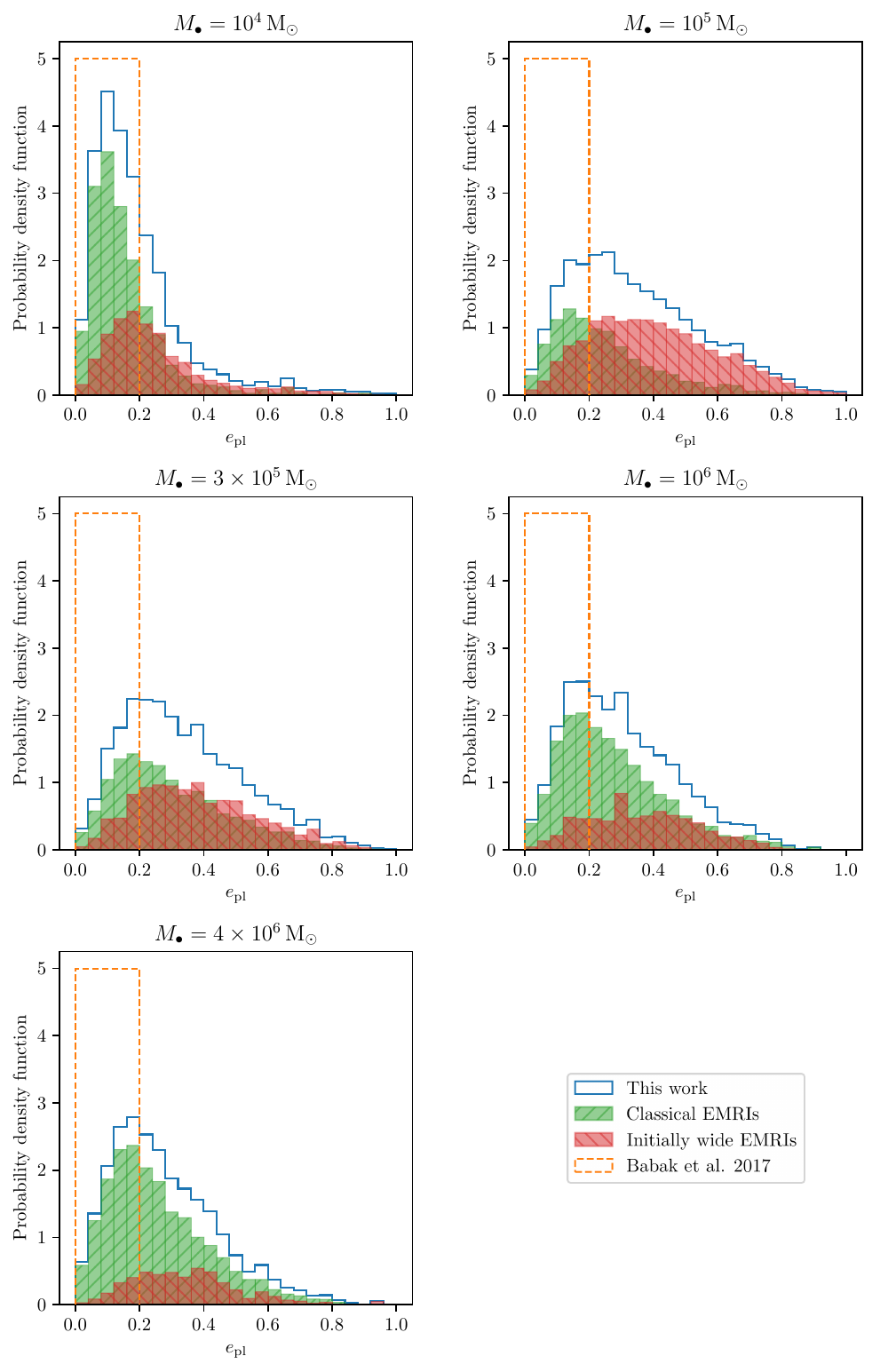}}
            \caption{EMRI eccentricity distribution at plunge, for different central MBH masses. Here we report the overall probability density function of our EMRI population (blue histograms) and display the underneath distributions of ``classical'' and ``initially wide'' EMRIs (green and red shaded histograms, see Sect.\ \ref{sec:results} for details), which are normalized to recover the overall distribution when summed together. We also show (orange dashed histogram) the distribution reported in \citet{2017PhRvD..95j3012B}, which has become a standard assumption in the literature.}
            \label{fig:plunge_distr}
    \end{figure*}

    We report the distributions of eccentricities at plunge $e_\mathrm{pl}$ for our EMRI populations in Fig.\ \ref{fig:plunge_distr}. We find that in the case of $M_\bullet = 10^4\, \mathrm{M_\odot}$ most of the EMRIs have $e_\mathrm{pl} \in (0,0.3)$ and the distribution is strongly peaked around $e_\mathrm{pl} \approx 0.1$, with a tail at larger eccentricities. For larger MBH masses the distribution shows a much broader peak around $e_\mathrm{pl} \approx0.2$, with a high fraction of EMRIs displaying eccentricities larger than 0.3. Indeed, for MBH masses of $10^5 \, \mathrm{M_\odot}$ and $3 \times 10^5 \, \mathrm{M_\odot}$, about 20 \% of EMRIs plunge at $e_\mathrm{pl} > 0.5$.

    We also separately display the distributions for ``classical EMRIs'' and ``initially wide EMRIs''. We make this distinction based on the initial $a_\mathrm{i}$ value used to initialize runs in \citet{2025A&A...694A.272M}. We classify the former (latter) as those such that $a_\mathrm{i} < a_\mathrm{c}$ ($a_\mathrm{i} > a_\mathrm{c}$). Here $a_\mathrm{c}$ is the critical semi-major axis at which the probability of forming EMRIs drops from 1 (at small $a$) to its value at large $a$. We define $a_\mathrm{c}$ such that
    \begin{equation}
        S(a_\mathrm{c}) = \frac{1}{2} \big( \min{S(a)} + \max{S(a)} \big) \, ,
    \end{equation}
    using the numerical $S(a)$ found in \citet{2025A&A...694A.272M} for each $M_\bullet$ case. The approximate values for $a_\mathrm{c}$ are \{$5\times10^{-3}, 9 \times10^{-3},10^{-2},10^{-2},10^{-2}$\} $R_\mathrm{inf}$ for $M_\bullet=$\{$10^4$, $10^5$, $3\times10^5$, $10^6$, $4\times10^6$\} $\mathrm{M}_\odot$ (see Fig.\ \ref{fig:veryniceplot}). We make this distinction because classically EMRIs are thought to only form for $a$ strictly less than $a_\mathrm{c}$, while a more thorough analysis shows that there is a smooth transition from the values of $a$ which allow for EMRI formation and those that do not, encompassing about an order of magnitude in $a$ \citep{2025A&A...694A.272M}. The wide EMRIs sub-population also includes (and for $M_\bullet \leq 3 \times 10^5 \, \mathrm{M_\odot}$ is mostly represented by) ``cliffhanger EMRIs'' \citep{2024PhRvL.133n1401Q,2025A&A...694A.272M}. As the flux of objects crossing into the loss cone typically peaks at $a > a_\mathrm{c}$ \citep{2022MNRAS.514.3270B}, these initially wide EMRIs can represent a significant fraction of the total. The only case where the flux peaks at $a < a_\mathrm{c}$ is that of $M=10^4 \, \mathrm{M_\odot}$. This is likely the reason why for this case only we observe a peak at lower eccentricity. In fact, the flux peaking at lower $a$ means that the binaries that are weighted more are those entering the GW-dominated region far away from the separatrix (see Fig.\ \ref{fig:veryniceplot}), and can thus circularize more before the final plunge.
    Figure \ref{fig:plunge_distr} shows that the two populations of EMRIs have different eccentricity distributions, with wide EMRIs being more eccentric when they plunge, with peaks at $e_\mathrm{pl}$ roughly double that of classical EMRIs. Despite this, the underlying populations are not separated enough for the overall distributions to show bi-modality. 

\begin{figure*}
        \centering
        \resizebox{\hsize}{!}{\includegraphics{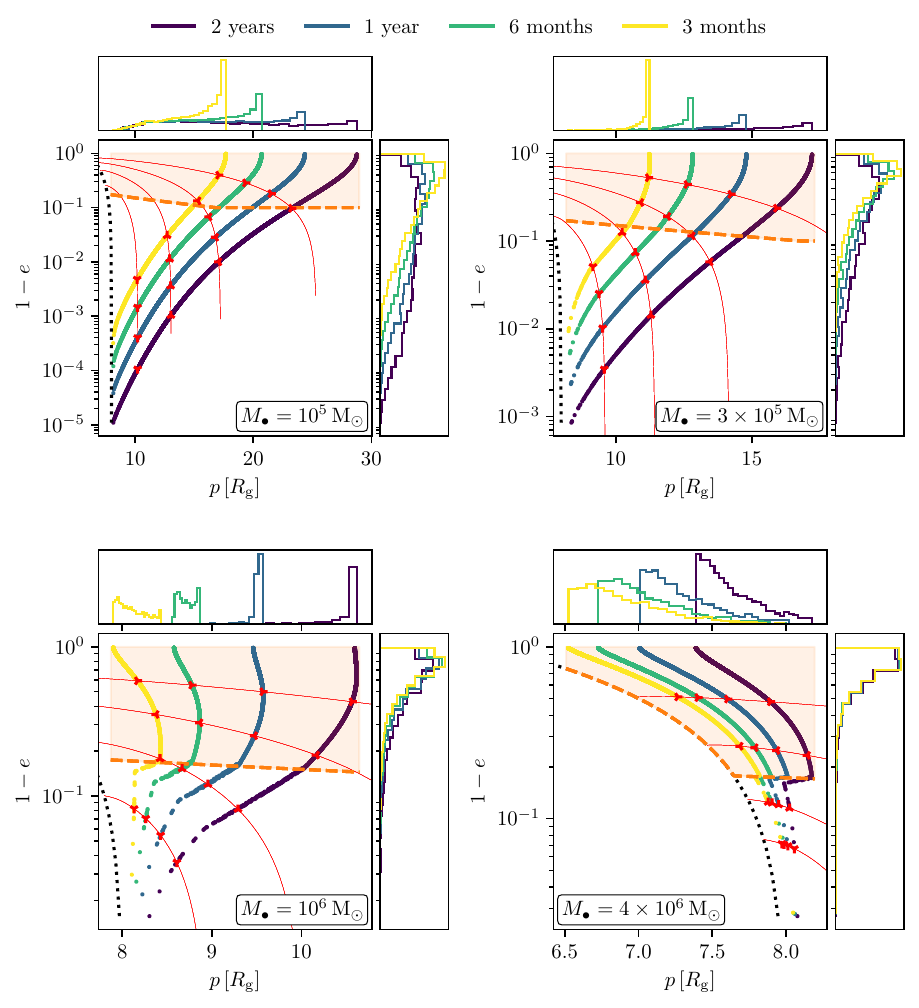}}
        \caption{EMRI $(p,e)$ locations 3 months, 6 months, 1 year and 2 years before the plunge, for different central MBH masses. Small panels at the top (right) of each large panel display the probability density function of semi-latus rectum (eccentricity). The orange dashed curve delimits the region (shaded in orange) where numerical fluxes in energy and angular momentum are currently available within FEW. Red markers and thin lines show the temporal evolution from our initial conditions $(p_0,e_0)$ of four selected binaries for each MBH mass, while the gray dotted curve is the separatrix.}
        \label{fig:times_distr}
    \end{figure*}

    In Fig.\ \ref{fig:times_distr}, we show the location on the $(p,e)$ plane of EMRIs at 3 months, 6 months, 1 year and 2 years before the plunge\footnote{The points are obtained by interpolating linearly between the two time-steps of the integration to plunge enclosing the target time.}. In some cases (especially for low $M_\bullet$) this required integrating backwards in time the final conditions of \citet{2025A&A...694A.272M}, and this led us to exclude $M_\bullet=10^4 \, \mathrm{M_\odot}$ from this analysis (see below). 
    The sharp transition that can be noticed for $M_\bullet = 10^6 \, \mathrm{M_\odot}$ and $M_\bullet = 4 \times 10^6 \, \mathrm{M_\odot}$ is due to binaries plunging outside the region where the \texttt{KerrEccEqFlux} trajectory module can be applied, as the \texttt{PN5} module overestimates the plunge time given the same initial conditions. We observe similar discontinuities even when switching between modules at later stages, much within the region for which numerical fluxes are available. We stress that the time to plunge is not an intrinsic physical property of the system, and there is no discontinuity in the evolution of the orbital parameters themselves, as shown both in Fig.\ \ref{fig:example} and by the thin red lines in Fig.\ \ref{fig:times_distr} itself.

    \begin{figure*}
        \centering
        \resizebox{\hsize}{!}{\includegraphics{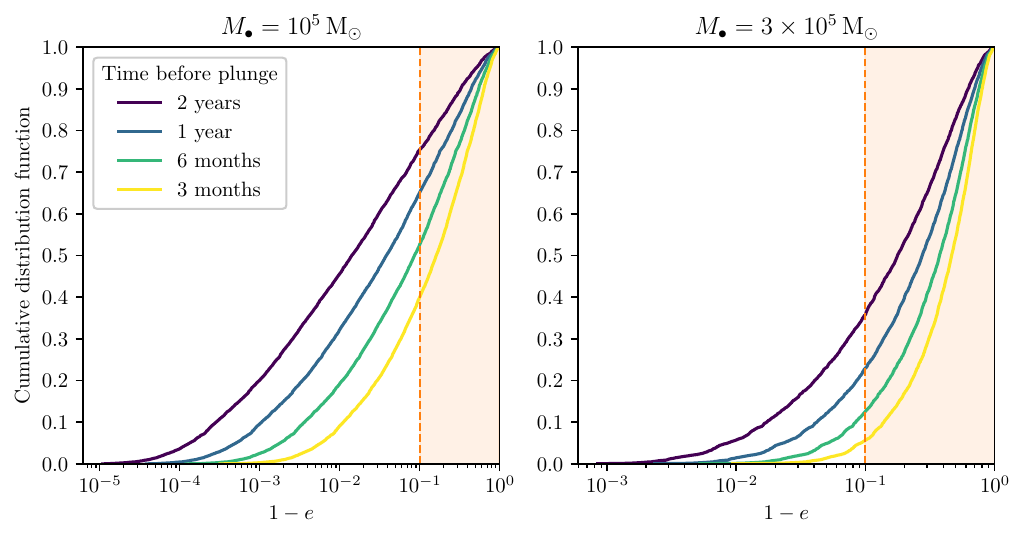}}
        \caption{EMRI eccentricity cumulative distribution function \{3 months, 6 months, 1 year and 2 year\} before the plunge, for $M_\bullet = 10^5 \, \mathrm{M_\odot}$ and $M_\bullet = 3 \times 10^5 \, \mathrm{M_\odot}$. The orange dashed line roughly delimits the region (shaded in orange) where numerical fluxes in energy and angular momentum are currently available within FEW. The region displayed here is valid only for $p \gtrsim 16.8 R_\mathrm{g}$, which is not always the case, so the fraction of EMRIs outside FEW range is slightly underestimated in the figure. See Table \ref{tab:percentages} for accurate values.}
        \label{fig:times2_distr}
    \end{figure*}

            \begin{table}
    \caption{\label{tab:percentages}Fraction of EMRIs outside the region where numerical fluxes in energy and angular momentum are currently available within FEW, for different MBH masses and times before the plunge.}
    \begin{ruledtabular}
    \begin{tabular}{ccccc}
     time & $10^5 \, \mathrm{M_\odot}$ & $3 \times 10^5 \, \mathrm{M_\odot}$ & $10^6 \, \mathrm{M_\odot}$ & $4 \times 10^6 \, \mathrm{M_\odot}$ \\ \midrule
            2 years & $75.4\,\%$ & $38.9\,\%$ & $5.3\,\%$ & $0.6\,\%$ \\
            1 year & $65.0\,\%$ & $27.8\,\%$ & $2.9\,\%$ & $0.5\,\%$ \\
            6 months & $52.5\,\%$ & $18.0\,\%$ & $1.6\,\%$ & $0.4\,\%$ \\
            3 months & $42.0\,\%$ & $11.0\,\%$ & $0.9\,\%$ & $0.4\,\%$ \\
    \end{tabular}
    \end{ruledtabular}
    \end{table}
    
    Figure \ref{fig:times_distr} shows that a large fraction of EMRIs, especially for low central MBH masses, falls outside of the current parameter space for which numerical fluxes are available within FEW. We better quantify this finding in Table \ref{tab:percentages}, which reports the fraction of EMRIs outside this domain as a function of MBH mass and time to plunge. Finally, a visual representation of this is provided in Fig.\ \ref{fig:times2_distr} for $M_\bullet = 10^5 \, \mathrm{M_\odot}$ and $M_\bullet = 3 \times 10^5 \, \mathrm{M_\odot}$. The cumulative distributions of eccentricity at the various time-to-plunge considered highlight that current EMRI waveforms cannot properly model up to 75\% of EMRIs with $10^5\,\mathrm{M_\odot}$ MBHs two years from plunge. On the one hand, this poses a challenge to early EMRI identification as well as proper parameter estimation; on the other hand, the lack of accurate waveforms will inevitably cause issues related to spurious residuals impacting the recovery of other sources within the LISA data global fit procedure.

    As mentioned above, in Fig.\ \ref{fig:times_distr}, we exclude the case $M_\bullet = 10^4 \, \mathrm{M_\odot}$ from the analysis as we find that a large fraction of systems spend much less than 2 years in the region of the orbital parameters where two-body relaxation is negligible (where $t_\mathrm{GW} < 10^{-3} \, t_\mathrm{rlx}$ in Fig.\ \ref{fig:veryniceplot}). This prevents trusting backward time integration from plunge for 2 years or more, as a significant fraction of systems was influenced by relaxation at that time. For $M_\bullet = 10^5 \, \mathrm{M_\odot}$, 13\% of binaries in our population stay in the region where $t_\mathrm{GW}<10^{-3} \, t_\mathrm{rlx}$ for less than 2 years and 3\% of them for less than 1 year. Binaries that plunge this quickly have small initial pericenter, which translates to large $e_0$ given that all of our binaries are initialized on the $t_\mathrm{GW}=10^{-3} \, t_\mathrm{rlx}$ curve shown in Fig.\ \ref{fig:veryniceplot}. In particular, for $M_\bullet = 10^5 \, \mathrm{M_\odot}$ binaries that take 1-2 years to merge have $1-e_0 \approx 4 \times 10^{-4}$, so that the distributions shown in Figs.\ \ref{fig:times_distr} and \ref{fig:times2_distr} might be increasingly less accurate for eccentricities larger than such value; we nevertheless expect two-body relaxation to still be subdominant compared to GWs emission. For $M_\bullet \geq 3 \times 10^5 \, \mathrm{M_\odot}$, there is no such issue.

    We also check if our binaries are within the LISA band by estimating their peak harmonic GW frequency $f_\mathrm{GW}^\mathrm{peak}$ during their inspiral, using the numerical fit from \citet{2021RNAAS...5..275H}:
    \begin{equation}
        f_\mathrm{GW}^\mathrm{peak} (p,e) = 2 \left( 1+\sum^4_{k=1} c_k e^k \right) (1-e^2)^{-3/2} f_\mathrm{orb}(p,e) \, ,
    \end{equation}
    where $c_1 = -1.01678$, $c_2=5.57372$, $c_3=-4.9271$, $c_4=1.68506$, and 
    \begin{equation}
        f_\mathrm{orb} (p,e) = \frac{1}{2\pi} \sqrt{\frac{GM_\bullet(1-e^2)^3}{p^3}}
    \end{equation}
    is the Keplerian orbital frequency. We find that the peak frequencies for all EMRIs in our population with $M_\bullet\geq 10^5$ stay between $10^{-4}$ to $1$ Hz during the whole integration. In the case of $M_\bullet=10^4 \, \mathrm{M_\odot}$ instead, EMRIs plunge at higher frequencies. All points plotted in Fig.\ \ref{fig:times_distr} have peak frequencies between $7 \times 10^{-4}$ and $7 \times 10^{-2}$ Hz.


\section{Comparison with previous works} \label{sec:comp}
    In this section we compare our findings about the eccentricity distribution with the few results quoted in the literature.

\begin{figure}
        \centering
        \resizebox{\hsize}{!}{\includegraphics{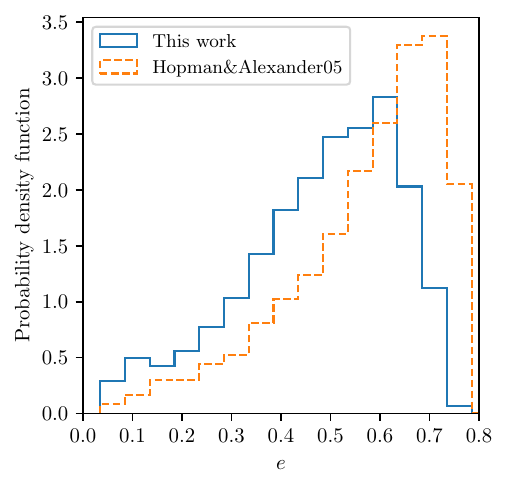}}
        \caption{EMRI eccentricity distribution for $M_\bullet=4\times10^6\,\mathrm{M_\odot}$ at $a = a_\mathrm{HA}$, compared to the results by \citet{2005ApJ...629..362H} with $M_\bullet=3\times10^6 \, \mathrm{M_\odot}$ at the same semi-major axis. A small fraction ($0.14 \%$) of EMRIs have already plunged before reaching $a_\mathrm{HA}$, and thus are not shown in this figure. They live in the small tail at $e_\mathrm{pl} \gtrsim 0.8$ in Fig.\ \ref{fig:plunge_distr}.}
        \label{fig:HA05}
    \end{figure}

    The first estimate of EMRI eccentricity can be found in 
    \citet{2005ApJ...629..362H}, who reported the eccentricity distribution for the case of $M_\bullet = 3 \times 10^6 \, \mathrm{M_\odot}$ at semi-major axis $a_\mathrm{HA} = 3.25\times10^{-6} \, \mathrm{pc}$ (about $22.6 \, R_\mathrm{g}$), so that the Keplerian period is $10^4 \, \mathrm{s}$, a threshold they used to estimate the entry in the LISA band. They performed Monte Carlo simulations, accounting for the effect of two-body relaxation and GWs emission on the energy and angular momentum of the orbits. Given a semi-major axis $a$, the maximum eccentricity a binary might have without plunging is \citep{2005ApJ...629..362H}
    \begin{equation}
        e_\mathrm{max} (a) = -\frac{R_\mathrm{g}}{a} +\sqrt{1-\frac{6R_\mathrm{g}}{a}+\frac{R_\mathrm{g}^2}{a^2}} \,
    \end{equation}
    which follows from Eq.\ \eqref{eq:sep} and $p=a(1-e^2)$. Consistently, their distribution (reported in Fig.\ \ref{fig:HA05}) shows a maximum eccentricity around $e_\mathrm{max}(a_\mathrm{HA}) = 0.81$. Despite not having performed simulations for $M_\bullet = 3 \times 10^6 \, \mathrm{M_\odot}$, we try to reproduce their results, using our data and weights for the $M_\bullet = 4 \times 10^6 \, \mathrm{M_\odot}$ case and evolving until the same semi-major axis\footnote{The eccentricity values are obtained by interpolating linearly between the two time-steps of the integration to plunge enclosing the condition $a(t)= a_\mathrm{HA}$.} $a_\mathrm{HA}$, finding a distribution slightly shifted to smaller values compared to theirs. Our main results extend their work to the plunge, and for a larger range of MBH masses, crucially including the contribution of wide and cliffhanger EMRIs.

    \citet{2021MNRAS.501.5012R} present eccentricity distributions for $M_\bullet = 4 \times 10^6 \, \mathrm{M_\odot}$, following an approach similar to that of \citet{2005ApJ...629..362H} and finding similar results. However, they terminate their simulations when the orbital period reaches $10^3 \, \mathrm{s}$, which corresponds to a final semi-major axis of $a_\mathrm{RP} = 7.71\times10^{-7} \, \mathrm{pc}$ (approximately $4.03 \, R_\mathrm{g}$). Since this radius lies within the Schwarschild separatrix for all values of $e$, we cannot perform a direct comparison of the eccentricity distribution.

     Distributions of $e_\mathrm{pl}$ spanning a range of MBH masses have been recently presented by \citet{2024ApJ...977....7R}.
     They first derive a steady-state stellar distribution surrounding the MBH by analytical means, assuming a population of stars of mass $m_\star = 1 \, \mathrm{M_\odot}$, and one of stellar-mass BHs of mass $m_\bullet = 10 \, \mathrm{M_\odot}$. In their work, stars and BHs are on profiles with inner slopes $1.5$ and $1.75$ respectively in the region of EMRI formation. In our simulations, the initial inner slope for both is $1.5$, but at the final stage the two components do not fit a power law profile. Nonetheless, at the mpc scale they are locally described by inner slope $1.3$ for stars and $1.7$ for BHs \citep{2022MNRAS.514.3270B}.
     
     They find an analytical form for the probability distribution function $P(e_\mathrm{pl})$ which ultimately depends only on the fraction of stellar-mass BHs over the total number of scatterers $f_\bullet$, and the mass ratio $m_\bullet / m_\star$. The mass ratio sets the critical fraction $f_\mathrm{c} = 4.5 \times 10^{-4} \big( (m_\bullet/m_\star)/10 \big)$ dividing between scarce ($f_\bullet < f_\mathrm{c}$) and plentiful ($f_\bullet > f_\mathrm{c}$) stellar-mass BH population.
    
    For $f_\bullet < f_\mathrm{c}$, they provide a long expression that we do not report, which has two branches\footnote{The $e<e_\mathrm{I}$ branch presents a typo: it should read $g(e_\mathrm{I})^{-9/2}$. The function $p(e)_{|f_\bullet <f_\mathrm{c}}$ as reported in their paper is not continuous at $e=e_\mathrm{I}$ and does not sum to 1.}, for $e < e_\mathrm{I}$ and $e>e_\mathrm{I}$, where $e_\mathrm{I}$ is determined by a function of $f_\mathrm{c}/f_\bullet$. In this case, they find that $P(e_\mathrm{pl})$ behaves like $e_\mathrm{pl}^{5/19}$ for $e_\mathrm{pl} \ll 1$, and like $e_\mathrm{pl}^{-49/19}$ for $e_\mathrm{pl} \gtrsim e_\mathrm{I}$. $P(e_\mathrm{pl})$ has a moving peak, but it never exceeds $e_\mathrm{pl} \approx 0.08$.
    
    For $f_\bullet > f_\mathrm{c}$, they report that the distribution stabilizes on\footnote{The result was first derived by \citet{2023ApJ...945...86L} in the context of stellar EMRIs.} $P(e_\mathrm{pl})= 2 g(e_\mathrm{pl})g'(e_\mathrm{pl})$, now being independent on the value of $f_\bullet$, where
    \begin{equation}
        g(e) = \frac{r_\mathrm{p}(e)}{r_\mathrm{p}(e=1)} = \frac{2e^{12/19}}{1+e} \left( \frac{304+121e^2}{425} \right)^{870/2299}
    \end{equation}
    and $g'(e)$ is its derivative with respect to $e$. Here they make use of \citep{1964PhRv..136.1224P}
    \begin{equation}
        r_\mathrm{p}(e) = (1-e)\,a(e) = \frac{c_0e^{12/19}}{1+e}\left( 1+\frac{121e^2}{304}\right)^{870/2299} \, ,
    \end{equation}
    where $c_0$ is determined by initial conditions, but its value is not important as it gets canceled out in computing $g(e)$. Thus, in this case they predict a distribution independent on $f_\bullet$ that peaks around $e_\mathrm{pl} \approx 0.08$, scales as $e_\mathrm{pl}^{5/19}$ at very low eccentricity, and then decays slowly.

    Although the general trends of their distribution are qualitatively comparable to ours, 
    they feature a peak at smaller $e_\mathrm{pl}$ and a slower decay as $e_\mathrm{pl}\rightarrow 1$ compared to what we observe. Even for $M_\bullet = 10^4 \, \mathrm{M_\odot}$, for which the location of the peaks is compatible, our decay at large eccentricity is better fitted by an exponential as opposed to a power law. The incompatibility is also caused by the fact that their $P(e_\mathrm{pl})$ does not smoothly go to $0$ when $e_\mathrm{pl}\to1$, and instead is sharply cut at some maximum value. For example, their model predicts that (independently on the MBH mass) $1 \%$ of EMRIs plunge at $e_\mathrm{pl}>0.9$ if $f_\bullet = 10^{-4}$, and $7 \%$ do the same if $f_\bullet > f_\mathrm{c}$. For comparison, we get between $0.1 \%$ and $0.5 \%$ depending on $M_\bullet$ for the same computation.
    
    While a precise one-to-one comparison between the two calculations is difficult, the observed differences may be due to their assumption that the system reaches a steady state. In fact, we find that for $M_\bullet <10^6 \, \mathrm{M_\odot}$ two-body relaxation continuously modifies the stellar distribution, so that both the EMRI formation rate and the stellar profiles are not steady (see Fig.\ \ref{fig:new_fig} and the upper panel of Fig.~\ref{fig:rates}). On the other hand, for $M_\bullet \geq 10^6 \, \mathrm{M_\odot}$ it takes more than $10$ Gyr for the onset of a profile determined entirely by relaxation, as this is the time required to lose memory of initial conditions inside $R_\mathrm{inf}$ \citep{2022MNRAS.514.3270B}.

    Around the same time our work was released, \citet{2025arXiv250900469S} reported EMRI eccentricity distributions for several cases, including the two-body formation channel. They evolve the energy and angular momentum distributions of stars and stellar-mass BHs in nuclear star clusters using a Fokker–Planck formalism, identify the boundary where $t_\mathrm{rlx} = t_\mathrm{GW}$, and transport the flux crossing it to orbits with pericenter $r_\mathrm{p} = 10 \, R_\mathrm{g}$. The resulting eccentricity distributions are consistently weighted through this flux, converting from energy and angular momentum to orbital parameters. Compared to our approach, theirs is inherently Newtonian and uses the GW emission equations from \citet{1964PhRv..136.1224P}. They also cannot account for cliffhanger EMRIs, as their method relies on orbit-averaged relaxation in the cluster.

\begin{figure}
        \centering
        \resizebox{\hsize}{!}{\includegraphics{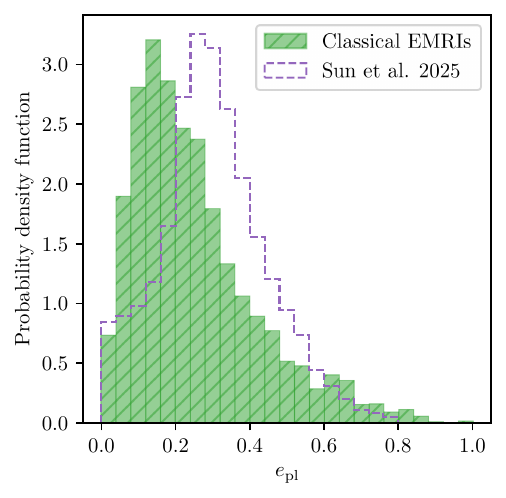}}
        \caption{EMRI eccentricity distribution at plunge for $M_\bullet=10^5\,\mathrm{M_\odot}$. The purple dashed histogram shows the time-averaged distribution from \citet{2025arXiv250900469S}, averaged up to $t=10 \, t_\mathrm{peak}$ and obtained under assumptions similar to ours (see text). The green histogram shows our results for the classical EMRI population only. The data are the same as in Fig.\ \ref{fig:plunge_distr}, with the normalization adjusted accordingly.}
        \label{fig:sun}
    \end{figure}

    The authors acknowledge some differences with respect to our work: \emph{(i)} they report eccentricities at $r_\mathrm{p} = 10 \, R_\mathrm{g}$ rather than at the separatrix; \emph{(ii)} they set $t_\mathrm{GW}$ to the approximate coalescence time from \citet{2020MNRAS.495.2321Z}, rather than the timescale over which GWs instantaneously impact the orbital elements that we adopt in \citet{2025A&A...694A.272M}\footnote{We remark that the relevant timescale for determining whether angular momentum relaxation dominates over GW emission is the time needed by GWs to impact the current orbital elements, which is non-trivially related to the time it takes for the binary to merge.}; \emph{(iii)} they derive orbital parameters from the Newtonian potential of the MBH plus the stellar potential, while we use the map from \citet{2024PhRvD.109f4077H}. In Appendix D of their work, all three differences are removed, yielding distributions qualitatively similar to ours (see their Fig.\ 32 and our Fig.\ \ref{fig:plunge_distr}), but missing a large fraction of high eccentricity events, consistent with their non-inclusion of initially wide EMRIs. Once those EMRIs are removed from our dataset, the distributions become much more similar in shape, although theirs peaks at slightly higher eccentricity (see Fig.\ \ref{fig:sun}). This is expected, as Peters' equations are known to overestimate eccentricity near plunge, where their underlying assumptions break down \citep{2021PhRvD.104j4023T,2025PhRvD.112b4012F}.

    Since a significant fraction of our initial conditions have $r_\mathrm{p} < 10 \, R_\mathrm{g}$, the direct comparison of the eccentricity distribution at $r_\mathrm{p} = 10 \, R_\mathrm{g}$ is practically unfeasible.

    Finally, \citet{2017PhRvD..95j3012B} reported a flat eccentricity distribution at plunge in the range $0 < e_\mathrm{pl}<0.2$, with a small tail of outliers with higher $e_\mathrm{pl}$. This distribution was derived as a suitable proxy to a set of numerical integrations of  EMRIs within the classical capture radius $a_\mathrm{c}$ to plunge. It should be noted, however, that although the $e_\mathrm{pl}$ information is given in their paper, it is not used at any point in their calculations, which focus on the computation of cosmological EMRI rates rather than on the determination of detailed EMRI orbital properties. Nevertheless, a flat $e_\mathrm{pl}$ distribution in the range $[0,0.2]$ has become a somewhat standard assumption when simulating EMRI populations. Our work demonstrates how inaccurate this assumption is (see Fig.~\ref{fig:plunge_distr}), and highlights the much larger eccentricity range covered by EMRIs formed via two-body relaxation.

\section{Conclusions} \label{sec:conc}
    In this work we derived realistic EMRI eccentricity distributions around Schwarzschild MBHs assuming the two-body relaxation channel. We quote the distribution at different stages of the inspiral, up to the final plunge. We accomplished this by integrating further the simulations performed by \citet{2025A&A...694A.272M} using the FEW package, appropriately weighting each run proportionally to the EMRI formation rate at its initial semi-major axis.

    We found that EMRIs can be highly eccentric at plunge, peaking around $e_\mathrm{pl} \approx 0.2$, but with a significant fraction at larger values, with up to $20 \%$ of EMRIs plunging at $e_\mathrm{pl}>0.5$ for some MHB masses. We also found that EMRIs forming at large semi-major axis, usually neglected in literature, tend to be (unsurprisingly) more eccentric at plunge. Moreover, we showed the expected location of EMRIs in the $(p,e)$ plane at different stages of the inspiral, highlighting how the current extent of numerical fluxes in energy and angular momentum available within FEW might not be able to cover the entire parameter space which is required for lower MBH masses. In the worst case scenario of $M_\bullet = 10^5 \, \mathrm{M_\odot}$, about $75 \%$ of EMRIs are outside the range of numerical fluxes at 2 years from plunge, and around $50 \%$ still are at 6 months from plunge.

    There are some effects that we neglected in this work, which could influence the eccentricity distribution of EMRIs. First, we only considered Schwarzschild BHs, whereas astrophysical BHs are generally expected to be spinning. The spin of the stellar-mass BH is unlikely to induce a significant deviation in the eccentricity of EMRIs at the level of precision considered here (the effect is small and enters at post-adiabatic order; see \emph{e.g.}\ Ref.\ \citep{2024PhRvD.109l4048B}). This, however, is likely not the case for the spin of the MBH. The loss of orbital energy and angular momentum differs for inspirals around Kerr MBHs, and, most importantly, the location of the separatrix depends on both the spin of the MBH and the inclination of the orbital plane with respect to the MBH spin axis \citep{2020PhRvD.101f4007S}. From an astrophysical perspective, this affects EMRI formation rates: plunging directly into a highly spinning MBH is harder on prograde orbits and easier on retrograde ones, compared to the Schwarzschild case (spin effectively shrinks or enlarges the loss cone region). When considering a population of EMRIs with generic inclinations around a Kerr MBH, the overall effect is expected to be an enhancement in the EMRI formation rate \citep{2013MNRAS.429.3155A}. We expect this effect to be more pronounced for initially wide orbits, thus increasing the number of highly eccentric EMRIs compared to the results presented here. We plan to properly investigate the impact of MBH spin in future work.
    
    Another simplification adopted in this study is the assumption that all stellar-mass BHs have the same mass of $10 \, \mathrm{M_\odot}$. A more realistic eccentricity distribution should account for a mass spectrum of the secondary objects and the corresponding mass segregation effects as the cluster evolves. We believe, however, that the impact of this would be small, since a continuous mass spectrum introduces a scatter in the slope of the distribution of the orbiting bodies comparable to the scatter observed during profile relaxation (compare our Fig.\ \ref{fig:new_fig} to Fig.\ 10 in Ref.\ \citep{2024ApJ...961..232Z}; our nuclear star cluster model has slope indices for stars and stellar-mass BHs close to their model M7). As discussed in Sect.\ \ref{sec:rates}, this scatter does not strongly affect the eccentricity distribution. Even if segregation might have a minor impact, heavier objects might have different eccentricity distributions because of stronger GW emission. This should be investigated further, especially as evidence for a subpopulation of stellar-mass BHs around $35 \, \mathrm{M_\odot}$ is emerging from GW observations \citep{2025arXiv250818083T}. Nonetheless, \citet{2005ApJ...629..362H} showed that the eccentricity distribution is not strongly affected if the secondary is assumed to be a lighter compact object, such as a neutron star or a white dwarf, and we might expect a similar behavior for more massive objects.

    Our simplified nuclear star cluster model also neglects the presence of binaries. Strong scatterings with tight binaries can lead to additional energy exchange due to the internal binding energy of the binary, influencing the relaxation process. As the average separation between objects in the nuclear star clusters we consider is much larger than typical binary separations ($0.1 \,\text{-}\, 100 \, \mathrm{AU}$), the exchange of energy with the inner binary might be subleading. Nevertheless, this possibility should be explored in future work to confirm that the eccentricity distributions inferred here are robust.

    Despite these caveats, our results call for an extension of the capabilities of FEW in order not to miss a large part of the EMRI population. We stress that any waveform model that aims at modeling astrophysical EMRIs has to cover a much wider parameter space that currently available. Accurate modeling of EMRIs is fundamental for parameter estimation, tests of general relativity and environmental effects, and for avoiding spurious residual in LISA global fit procedures. Precisely detecting high eccentricity EMRIs is also essential if we wish to infer the underling formation rate of each channel, which in turn will reveal more about the nature of MBHs and their host galaxies.

\begin{acknowledgments}
We thank Niels Warburton for useful discussions during the ``Enabling Future Gravitational Wave Astrophysics in the Milli-Hertz Regime'' workshop, as well as the organizers and administrative staff for ensuring a stimulating and well-run meeting. We also thank Monica Colpi for her valuable comments and suggestions.
\\
D.M.\ acknowledges that this publication was produced while attending the PhD program in Space Science and Technology at the University of Trento, Cycle XXXIX, with the support of a scholarship financed by the Ministerial Decree no.\ 118 of 2nd March 2023, based on the NRRP - funded by the European Union - NextGenerationEU - Mission 4 ``Education and Research'', Component 1 ``Enhancement of the offer of educational services: from nurseries to universities'' - Investment 4.1 ``Extension of the number of research doctorates and innovative doctorates for public administration and cultural heritage'' - CUP E66E23000110001.
A.S.\ acknowledges the financial support provided under the European Union’s H2020 ERC Advanced Grant ``PINGU'' (Grant Agreement: 101142079).
M.B.\ acknowledges support from the Italian Ministry for Universities and Research (MUR) program “Dipartimenti di Eccellenza 2023-2027”, within the framework of the activities of the Centro Bicocca di Cosmologia Quantitativa (BiCoQ). This research was supported by the Munich Institute for Astro-, Particle and BioPhysics (MIAPbP) which is funded by the Deutsche Forschungsgemeinschaft (DFG, German Research Foundation) under Germany´s Excellence Strategy – EXC-2094 – 390783311.
\end{acknowledgments}

\section*{Data availability}
Some of the data that support the findings of Fig.\ \ref{fig:plunge_distr} are openly available on Zenodo \citep{dataZenodo}. The data are available upon reasonable request from the authors.

\bibliographystyle{apsrev4-2-author-truncate-article-titles}
\bibliography{bibliography}

\begin{appendix}

\section{Alternative weighting} \label{app:alternative}
    In this appendix, we examine how the results in Fig.\ \ref{fig:plunge_distr} change when using an alternative weighting procedure based on the classical criterion for EMRI formation. In the standard approach \citep{2005ApJ...629..362H}, BHs entering the loss cone are classified as EMRIs if $a < a_\mathrm{c}$, and as direct plunges if $a > a_\mathrm{c}$. By contrast, in Sect.\ \ref{sec:rates} we employed a transfer function $S$ to provide a smoother transition between the two regimes and to account for cliffhanger EMRIs.

    In the alternative procedure, we replace $S$ with a step function to separate EMRIs from direct plunges. To do this, we determine the critical energy $\mathcal{E}_\mathrm{c}$ such that
    \begin{equation}
        S(\mathcal{E}_\mathrm{c}) = \frac{1}{2} \big( \min{S(\mathcal{E})} + \max{S(\mathcal{E})} \big) \, ,
    \end{equation}
    in the energy range explored in \citet{2025A&A...694A.272M}. We then re-process the same simulations as in Sect.\ \ref{sec:rates} to compute the formation rate of EMRIs according to the classical differential rate
    \begin{equation}
        \mathcal{F}^\mathrm{cl}_\mathrm{EMRI} = \mathcal{F}_\mathrm{lc}(\mathcal{E}) \, \Theta(\mathcal{E}-\mathcal{E}_\mathrm{c})
    \end{equation}
    where $\Theta$ is the Heaviside step function. The weights are then computed in the same way as described in Sect.\ \ref{sec:rates}.

\begin{figure*}
        \centering
        \resizebox{0.775\hsize}{!}{\includegraphics{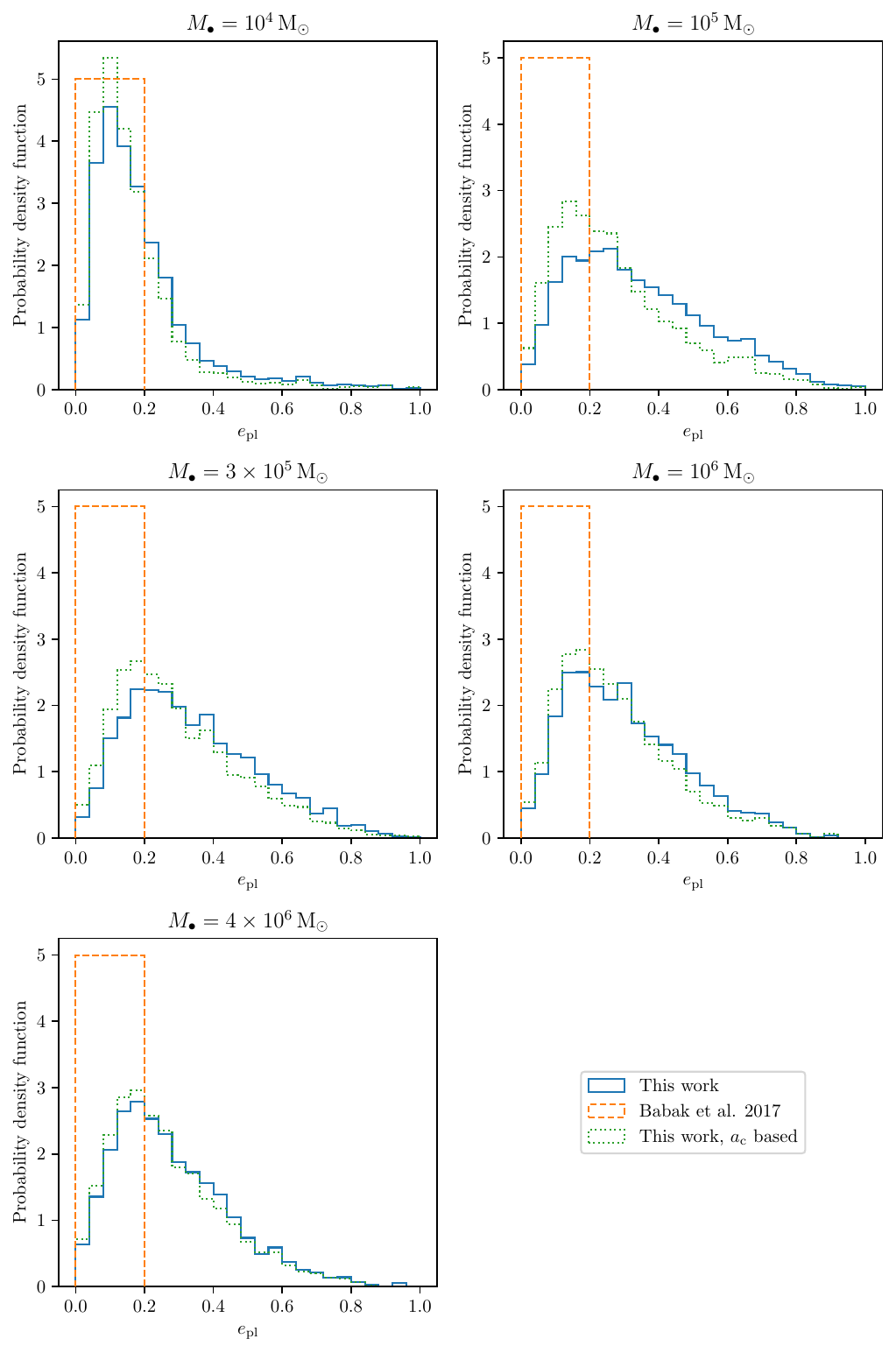}}
        \caption{EMRI eccentricity distribution at plunge, for different central MBH masses. Here we report the probability density function of our EMRI population analyzed with the procedure described in Sect.\ \ref{sec:rates} (blue solid histograms) and the one described in this appendix (green dotted histograms). We also show (orange dashed histogram) the distribution reported in \citet{2017PhRvD..95j3012B}, which has become a standard assumption in the literature.}
        \label{fig:plunge_distr_ac}
    \end{figure*}

    The resulting eccentricity distributions at plunge are shown in Fig.\ \ref{fig:plunge_distr_ac}. As expected, they resemble the classical EMRI distributions of Fig.\ \ref{fig:plunge_distr} with more pronounced peaks at lower $e_\mathrm{pl}$, since wide EMRIs are excluded. The match is not exact, however, because classical EMRIs with $a_\mathrm{i} \approx a_\mathrm{c}$ receive a higher weight.

    The $10^5 \, \mathrm{M_\odot}$ and $3 \times 10^5 \, \mathrm{M_\odot}$ MBH cases are most affected by this change in weighting, as cliffhanger EMRIs form in these regimes. Figure \ref{fig:plunge_distr_ac} illustrates how, in these cases, the common view of EMRI formation underestimates the fraction of highly eccentric EMRIs at plunge.
    
\end{appendix}

\end{document}